%% file: Maintexts.tex
\documentclass[%
 reprint,
showpacs,
superscriptaddress,
prl,
amsmath,amssymb,
aps,
]{revtex4-1}

\usepackage{graphicx}
\usepackage{dcolumn}
\usepackage{bm}
\usepackage{hyperref}
\usepackage{color}
\newcommand{\revisionHb}[1]{\textcolor{black}{{#1}}}
\newcommand{\revisionHa}[1]{\textcolor{black}{{#1}}}

\begin{document}


\title{\revisionHb{Probing breakdown of topological protection:   Filling-factor-dependent evolution of robust quantum Hall incompressible phases}}

\author{T. Tomimatsu}
\affiliation{%
Graduate School of Sciences, Tohoku University, Sendai 980-8578, Japan
}%

\author{K. Hashimoto}%
\email{hashi@tohoku.ac.jp}
\affiliation{%
Graduate School of Sciences, Tohoku University, Sendai 980-8578, Japan
}
\affiliation{%
Centre for Spintronics Research Network, Tohoku University, Sendai 980-8578, Japan}
\author{S. Taninaka}
\affiliation{%
Graduate School of Sciences, Tohoku University, Sendai 980-8578, Japan
}%
\author{S. Nomura}
\affiliation{%
Division of Physics, University of Tsukuba, Tennodai, Tsukuba 305-8571, Japan
}%
\author{Y. Hirayama}%
\affiliation{%
Graduate School of Sciences, Tohoku University, Sendai 980-8578, Japan
}%
\affiliation{%
Centre for Spintronics Research Network, Tohoku University, Sendai 980-8578, Japan}
\affiliation{%
Center for Science and Innovation in Spintronics (Core Research Cluster), Tohoku University, Sendai 980-8577, Japan}

\date{\today}

\begin{abstract}
\revisionHa{The integer quantum Hall (QH) effects characterized 
by topologically quantized and nondissipative transport
are caused by an electrically insulating incompressible phase
that prevents backscattering between chiral metallic channels.}
\revisionHb{We probed the incompressible area susceptible to the breakdown of  topological protection using a scanning gate technique incorporating nonequilibrium transport.
The obtained pattern revealed
the filling-factor ($\nu$)-dependent
evolution of the microscopic incompressible structures 
located along the edge and in the bulk region. 
We found that these specific structures, respectively 
attributed to the incompressible edge strip
and bulk localization, show good agreement in terms of $\nu$-dependent evolution
with a calculation of the equilibrium QH incompressible phases,
indicating the robustness of the QH incompressible phases 
under the nonequilibrium condition. } 
Further, we found that the $\nu$ dependency of the incompressible patterns is, 
in turn, destroyed by a large imposed current 
during the deep QH effect breakdown.
These results demonstrate the ability of our method to image 
the microscopic transport properties of a topological two-dimensional system.

\end{abstract}

\pacs{Valid PACS appear here}
\maketitle

A two-dimensional electron system (2DES) 
subjected to strong magnetic fields forms 
a quantum Hall (QH) insulating phase 
with a state lying in a gap 
between quantized Landau levels (LLs).
This gapped \revisionHa{phase}, 
the so-called incompressible \revisionHa{phase},
prevents backscattering between 
the metallic gapless (compressible) \revisionHa{phase} 
counter-propagating along both sides of the 2DES edges
\cite{chakraborty2013QHtextbook}.
This is the key microscopic aspect of 
nondissipative chiral transport of the integer QH effect, 
which is characterized by 
\revisionHa{a longitudinal resistance that vanishes and a}
universal quantized Hall conductance
protected by a topological invariant
\cite{TopologicalQHThouless1982,TopologicalQHHatsugai1993}.
Topological phases are attracting renewed attention 
due to the recent discovery of exotic topological materials 
such as insulators 
\cite{TopoInsulatorKane2005,TopoInsulator2Kane2005,QSHEprediction2006,QAHEexperiment2013}, 
superconductors \cite{superMayoranaTheorem2008}, 
and Weyl semimetals \cite{weyl2011}.

The formation of incompressible and compressible \revisionHa{phases} 
in the QH regime originates from the interplay between Landau quantization 
and the Coulomb interaction \cite{CSG1992}, 
which drives nonlinear screening \cite{wulfGerhardts1988nonlinear,efros1993nonlinear}. 
\revisionHb{The} spatial configuration depends 
on the potential landscape.
\revisionHa{For example,} 
the edge confinement potential, 
accompanied by strong bending of the LLs,
forms spatially alternating unscreening and screening regions
due to the Fermi-level pinning at the gap and LLs.
These regions respectively result in 
alternating incompressible and compressible strips 
near the 2DES edge.
The innermost incompressible strip moves 
and spreads to the bulk 
as the LL filling factor $\nu$ reduces 
to an \revisionHb{integer $i$} from a higher $\nu$.
\revisionHa{This $\nu$ dependency of} 
the edge strips has been microscopically 
investigated using 
\revisionHa{various} imaging techniques such as
\revisionHa{single-electron transistor imaging \cite{YacobySET},} 
Hall-potential imaging \cite{AhlswedeWeis2001Hallpotential,Weis2011},
microwave impedance imaging \cite{lai2011MicrowaveImpedance}, 
capacitance imaging \cite{suddards2012capacitance},
and scanning gate imaging \cite{SGMwoodside,SGMIhnQHE,paradiso2012QPCprl,pascherIhn2014QPC,GhrapheneSGM,SGM_HgTe,SGMgrapheneDirac}, \revisionHa{and it has been extended 
with superconducting quantum interference device
(SQUID) magnetometry \cite{SQUIDspinHall2013}
for} a topological spin-Hall insulator.

In the bulk incompressible region 
\revisionHa{formed at $\nu\simeq i$}, 
the disorder potential plays an important role 
in giving rise to isolated compressible puddles 
that result in QH localization \cite{LocalizationReview,InteractionTheoryIllani,RudoNewJPhys2007}.
\revisionHa{QH localized states have been probed} \cite{ilani2004SETprobe,Ashioori2005Capacitance},
and they were demonstrated 
to undergo phase transition to a delocalized state
with a scanning tunneling microscope \cite{Hashi2008QHT}, 
which accounts for the transition from nondissipative 
to dissipative transport \cite{LocalizationReview,InteractionTheoryIllani,RudoNewJPhys2007}.

\revisionHb{By contrast,} 
for a practical sample such as a Hall bar, 
microscopic pictures of the QH effect
\revisionHb{have} not \revisionHb{been} fully understood \cite{Klitzing2019essay}.
For instance, \revisionHb{the contribution of the innermost 
incompressible strip to} nondissipative transport 
is \revisionHb{debatable}
\cite{SiddikiGerhardts2004}. 
\revisionHb{Moreover, the microscopic mechanism of the breakdown of topological protection
is thought to originate from  
backscattering through the incompressible region \cite{eaves1984LLscattering,eavesTurbulence},
and this has recently become a key issue in research on the quantum spin Hall effect
\cite{SGM_HgTe,QSHE_TMD}
and anomalous QH effect\cite{QAHEbackscattering}.}
To understand the transport properties inherent to the QH effect,
it is important to elucidate the microscopic aspects 
of QH transport \revisionHa{in real devices---namely, how 
the innermost incompressible region evolves in the Hall bar
\revisionHb{and, hence, limits non-dissipative, topologically quantized transport.}}
Here, we present the ability of a novel scanning-gate method
incorporating a nonequilibrium transport technique 
to  
\revisionHb{pinpoint the areas susceptible to breakdown 
of topological protection 
and hence gives access to the local breakdown 
at a nonequilibrium transport.
}
\revisionHb{We found the robust QH nature, i.e., the}
evolution of the \revisionHa{innermost} incompressible QH \revisionHa{phase} from the edge strip to bulk localization,
\revisionHb{and the microscopic aspects of the global QH breakdown
---namely, the breakdown of the QH effect, 
under a larger nonequilibrium condition}. 

To probe \revisionHb{the local breakdown of the incompressible area}, 
we used a powerful tool 
(the scanning gate microscope (SGM)), which 
\revisionHa{uses resistive detection 
as transport measurements and hence directly images 
transport channels with a high sensitivity 
at a high spatial resolution \cite{topinka2000SGMscience}.
In topological phases such as QH
\cite{SGMwoodside,aoki2005,SGMIhnQHE,paradiso2012QPCprl,pascherIhn2014QPC}, 
and quantum spin Hall \cite{SGM_HgTe}, 
the SGM has been widely used to probe
an important local transport characteristic,
namely backscattering 
between counter-propagating edge channels.
To capture a sufficiently intense signal from backscattering---one that is strongly suppressed \revisionHb{at $\nu\simeq i$}, 
a conventional SGM requires the electrostatic influence 
of a large negative tip voltage $V_{\mathrm{tip}}\leq-1$ V \cite{paradiso2012QPCprl,pascherIhn2014QPC,aoki2005}, 
moving the incompressible strip across a narrow channel.
However, the tip-induced displacement of the incompressible strip may locally disturb the intrinsic structure.}

\begin{figure}
\includegraphics[width=1.0\linewidth]{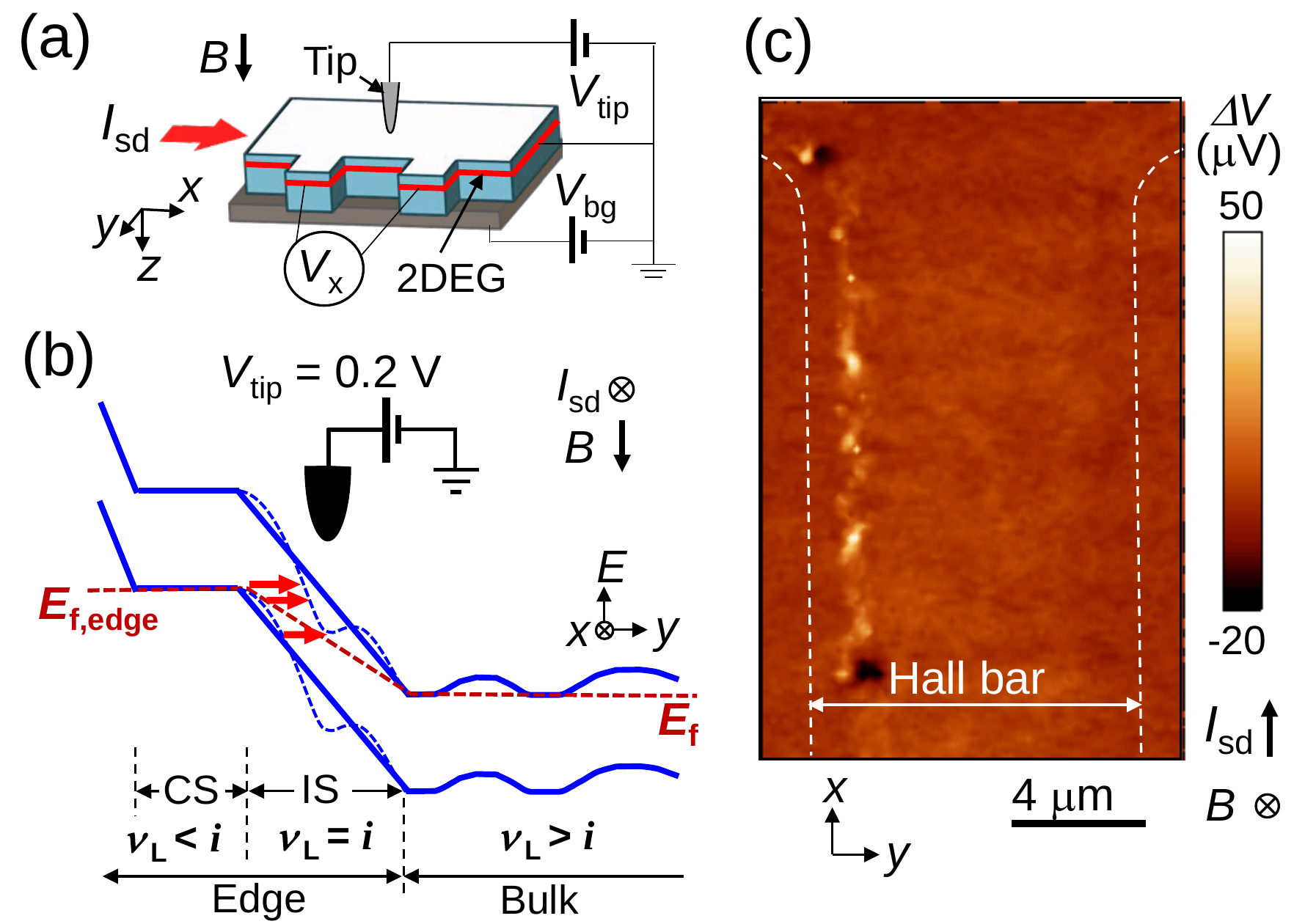}
\caption{(a) Schematic of the experimental setup of the SGM. 
$V_{\rm{x}}$ is recorded as the tip is scanned at $V_{\rm{tip}}$
with $I_{\rm{sd}}$ and $B$ respectively applied 
in the $x$ and $z$ directions. 
(b) Schematic of inter-LL tunneling (marked by red arrows)
between two LL subbands near the edge of 
the higher $\mu_{\rm{chem}}$ side
under the nonequilibrium condition, 
which is derived from the deviation 
between the Fermi energies in the edge ($E_{\rm{f,edge}}$) 
and bulk ($E_{\rm{f}}$) compressible regions.
\revisionHa{Tip-induced LL bending,
indicated by the blue dashed line, enhances inter-LL tunneling.} 
The incompressible and compressible strips 
are indicated by ``IS'' and ``CS,'' respectively.
``Edge'' and ``Bulk'' indicate the edge strips 
and 2DES bulk region, respectively. 
\revisionHa{The magnitudes of $\nu_{\rm{L}}$ 
with respect to the exact integer \revisionHb{$i$} are also shown for each region.}
(c) SGM image of a tip-induced $V_{\rm{x}}$ change 
($\Delta V$) at $\nu =2.27$, $I_{\rm{sd}}= 3.1\ \mu$A, 
and $B = 4$ T; dashed lines denote the Hall bar edges.
The line noise was removed 
\revisionHa{using} 2D Fourier filtering.}
\label{SGM}
\end{figure}
\revisionHa{In our setup shown in Fig. \ref{SGM}(a),}
to minimize global perturbation, 
the tip voltage was set to $V_{\rm{tip}} \sim 0.2$ V, 
corresponding to the value of the contact potential mismatch
between the tip and the sample \cite{Supple}. 
To address the QH \revisionHa{phase} \revisionHb{at $\nu\simeq i$} 
and obtain a local signal without applying a large tip voltage,
we incorporated nonequilibrium transport.  
We investigated a 2DES that was confined 
in a 20-nm-wide GaAs/Al$_{0.3}$Ga$_{0.7}$As 
quantum well located 165 nm beneath the surface. 
The wafer was processed into a 10-$\mu$m-wide Hall bar.
The mobility of the 2DES was
$\mu=130 \ \rm{m^{2}V^{-1}s^{-1}}$
at an electron density 
$n = 1.8\times10^{15}\rm\ m^{-2}$.
Figure \ref{SGM}(b) depicts the alternating compressible 
and incompressible regions formed along an edge of the Hall bar.
The local $\nu$ ($\nu_{\rm{L}}$) 
of an incompressible strip is \revisionHa{maintained at} \revisionHb{$\nu_{\rm {L}}=i$}, 
while the bulk $\nu$ is modified by sweeping $B$ 
or $n_{\rm{s}}$ tuned by a back gate voltage ($V_{\rm{bg}}$).
To achieve the nonequilibrium condition,
we increased the source--drain current ($I_{\mathrm{sd}}$) 
until the Hall voltage deviated 
from the QH condition. 
The imposed Hall voltage
\revisionHa{predominantly enhances the potential slope 
within the innermost incompressible region \cite{Weis2011},
inducing} 
inter-LL tunneling from the edge to the bulk
through the innermost incompressible strip.
\revisionHb{Then, an electron is backscattered into the opposite edge channel
through compressible or directional-hopping channels 
along the Hall electric field.}
\revisionHa{This leads to a} dissipative current \cite{Eaves1986,panos2014SHM_QHBD},
and thus a nonzero longitudinal resistance. 
\revisionHb{ 
The tip locally provides small electric perturbation 
to 2DEG owing to the effective potential mismatch
rearranged by imposed excess Hall voltage,
and eventually bends the LL locally and, hence, increases 
the inter-LL tunneling rate,
as shown in Fig. \ref{SGM}(b) 
(details are discussed 
in the Supplementary Material \cite{Supple}).
The tip induced inter-LL tunneling
further enhances backscattering 
and hence the longitudinal voltage ($V_{\rm{x}}$),
By mapping the resulting $\Delta V$, we can} 
visualize the innermost incompressible \revisionHa{region}. 
\revisionHa{All measurements were performed at a sample temperature below 250 mK.}

Figure \ref{SGM}(c) shows a typical SGM image 
obtained by capturing $\Delta V$ at $\nu=2.27$ ($B = 4$ T) 
\revisionHa{under the nonequilibrium} conditions at $I_{\rm{sd}}=$ 3.1 $\mu$A.
A distinct line-like pattern can be seen 
extending in the $x$ direction along a Hall bar edge (left dashed line), 
which corresponds to the side 
with the higher chemical potential ($\mu_{\rm{chem}}$) 
across the $y$ direction of the Hall bar.
This $\mu_{\rm{chem}}$ dependency, 
confirmed by reversing the direction 
\revisionHa{of the current} \cite{Supple}, 
can be explained by the fact that 
$\mu_{\rm{chem}}$ mainly drops  
at the higher-$\mu_{\rm{chem}}$ incompressible strip 
in a nonequilibrium condition \cite{panos2014SHM_QHBD},
where \revisionHb{we expect} a higher rate of inter-LL scattering \cite{HashiSNRM}
and thus \revisionHb{more} SGM sensitivity 
with respect to the corresponding incompressible strip.
To minimize the influence of $I_{\rm{sd}}$ on the incompressible patterns, 
$I_{\rm{sd}}$ 
\revisionHa{was limited to below the current in all measurements,}
at which the position of the strip shows 
no significant $I_{\rm{sd}}$ dependence
in the entire measurement region of $\nu$.
Otherwise, there is a deviation associated with \revisionHb{global} QH breakdown, 
as discussed in the Supplementary Material \cite{Supple}. 
\begin{figure}
\includegraphics[width=1.0\linewidth]{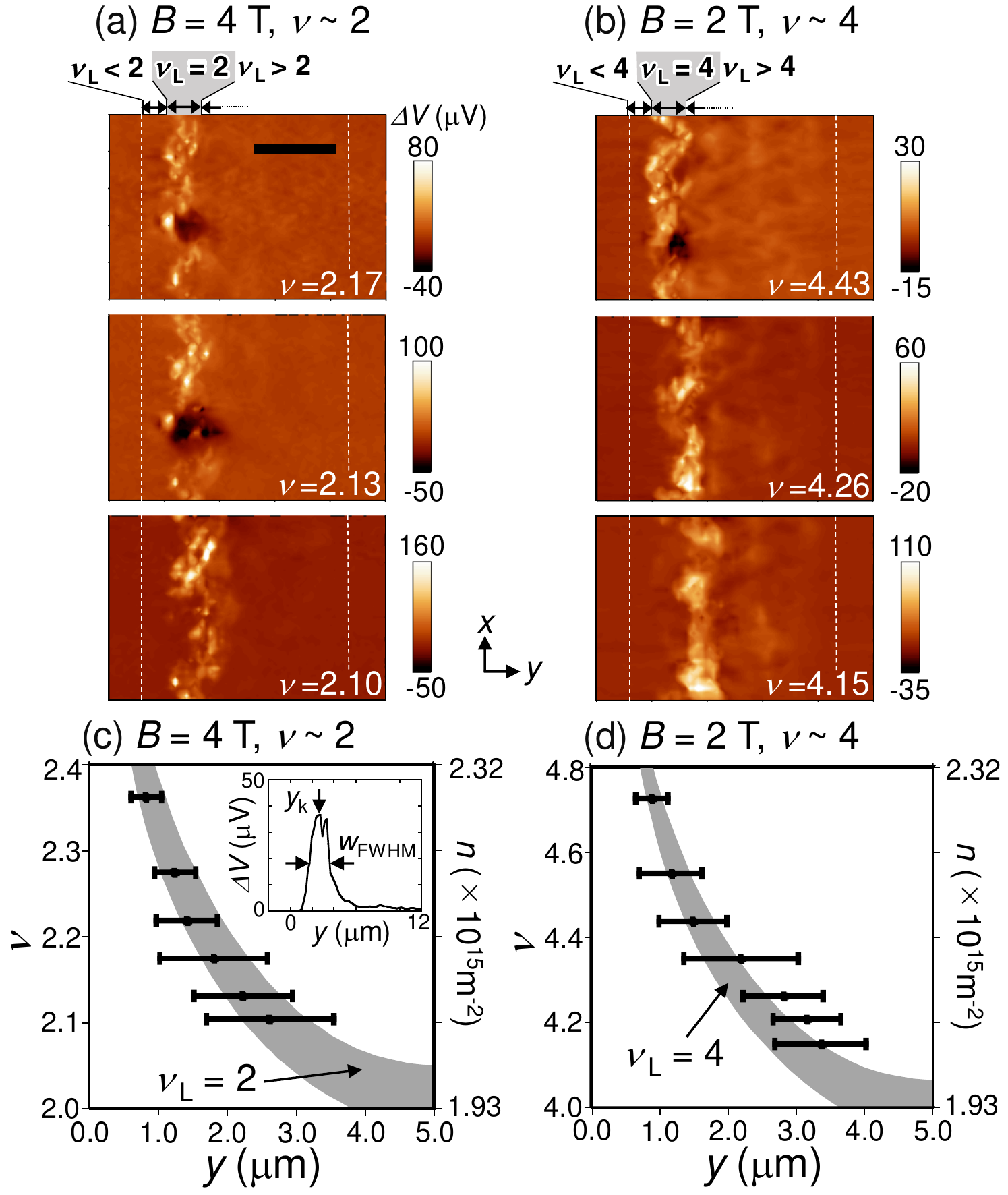}
\caption{SGM measurements near $\nu = 2$ and 4.
(a), (b) Representative $\Delta V$ images at different $\nu$ 
tuned by $n$ at (a) $B = 4$ T near $\nu=2$  
and (b) $B = 2$ T near $\nu=4$ 
\revisionHb{(see the Supplementary Material \cite{Supple} 
for the corresponding longitudinal and Hall resistances at equilibrium in the QH regime)}.
The scale bar is 4 $\mu$m.
Dashed lines denote the Hall bar edges.
The line noise was removed 
using 2D Fourier filtering.
As $\nu$ decreases toward integer \revisionHb{$i$},
the patterns are further enhanced, 
such that the full scale of contrast 
is appropriately optimized for clarity.  
(c), (d) Position (dot) and width (bar) 
of the $\Delta V$ peak as a function of  $\nu$ 
near (c) $\nu=2$  and (d) 4 .
The position and width are respectively determined
by the distance from the Hall bar edge ($y_{\rm{k}}$)
and the full width at half maximum ($W_{\rm{FWHM}}$).
These are extracted from the cross-sectional
$\Delta V$ profile spatially averaged over the $x$ region 
($\overline{\Delta V}$), as indicated in the inset in (c)
[obtained from the image for $\nu=2.10$ in (a)].
Here, $\nu=2.0$--2.4 and 4.0--4.8 are selected 
\revisionHa{to ensure} the same $n_{\rm{s}}$ range $1.93\times10^{15}$--$2.32\times10^{15} \rm{m^{-2}}$.
The gray area denotes the incompressible regions determined 
by LSDA calculation for (c) $\nu_{\rm{L}}$=2 and (d) 4.
}
\label{nu2map}
\end{figure}

\revisionHb{We} examined the $\nu$ dependence of the SGM patterns. 
The measurements were performed at $I_{\rm{sd}}$,
which was tuned to maintain \revisionHa{a constant} offset 
$V_{\rm{x}}$ ($V_{\rm{x}}\sim 1$ mV)
at each $\nu$ (for details regarding the conditions, 
see the Supplementary Material \cite{Supple}). 
Figure \ref{nu2map}(a) displays representative SGM images
captured near $\nu=2$ and tuned
with the gate-controlled $n_{\rm{s}}$ 
at constant $B$ ($B=4$ T).
Decreasing $\nu$ from 2.17, the position of the line pattern 
shifts and widens to the bulk of 2DES.
The same tendency of the $\nu$-dependent patterns
was also observed in the same area near $\nu=4$ 
at $B=2$ T, as shown in Fig. \ref{nu2map}(b).

We extracted the positions ($y_{\rm{k}}$) 
and width ($W_{\rm{FWHM}}$) of the line patterns, 
which were respectively defined as the first moment 
(for details, see the Supplementary Material \cite{Supple})
and the full width at half maximum in a $\Delta V$ profile 
after spatially averaging over the 8.5-$\mu$m range 
in the $x$ direction ($\overline{\Delta V}$), 
as shown in the inset of Fig. \ref{nu2map}(c). 
For a quantitative comparison of the observed $\Delta V$ peak positions, 
we performed a calculation in the Landau gauge based 
on the local spin-density approximation (LSDA) \cite{nomura2004optical,nomura2015NanoLett}
using a typical potential profile \cite{GuvenGerhardts2003} 
in the QH regime ($I_{\rm{sd}}=0$ A) (for details regarding the calculations, 
see the Supplementary Material \cite{Supple}). 
The $\nu$-dependent \revisionHa{incompressible region}, 
experimentally determined from $y_{\rm{k}}$ (dots) 
and $W_{\rm{FWHM}}$ (bars), 
is compared with
the innermost QH incompressible 
\revisionHa{region (gray area)} calculated by the LSDA 
for $\nu_{\rm{L}}=2$ in Fig. \ref{nu2map}(c)
and $\nu_{\rm{L}}=4$ in Fig. \ref{nu2map}(d).
We found good agreement between the experimental results 
and the LSDA calculation for both values of $\nu$. 
Additionally, a closer examination of the line pattern 
shows local fluctuation in the same region at both values of $\nu$,
e.g., in the bottom half of the images
taken at $\nu=2.17$ in Fig. \ref{nu2map}(a)
and at $\nu=4.43$ in Fig. \ref{nu2map}(b).
This implies that the edge incompressible strip
meanders along the equipotential line 
disturbed by potential disorder.

\begin{figure}
\includegraphics[width=0.9\linewidth]{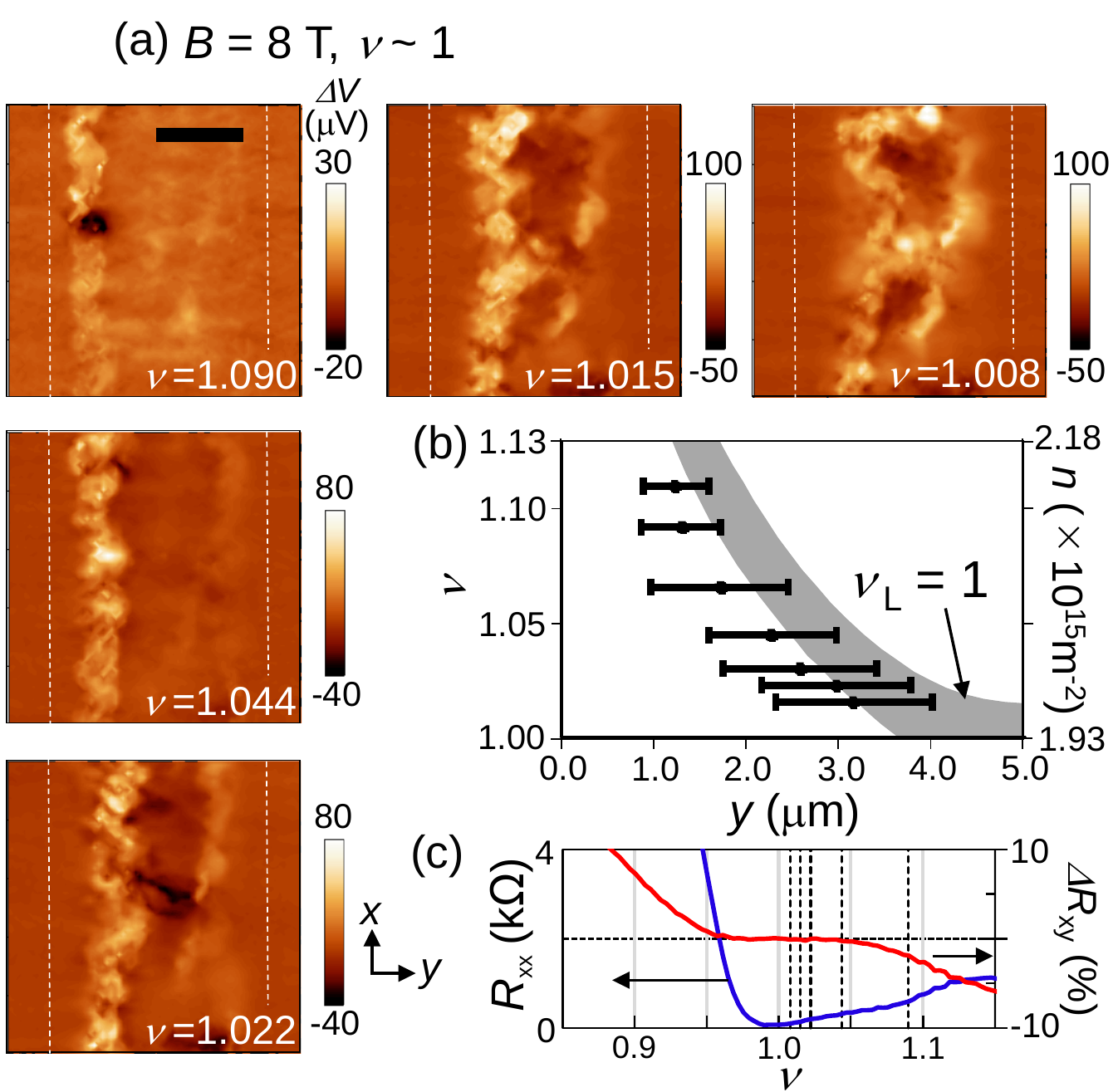}
\caption{SGM measurements near $\nu = 1$.
(a) Representative $\Delta V$ images taken
at different $\nu$ tuned by $n$ at $B = 8$ T.
The scale bar is 4 $\mu$m.
Dashed lines denote the Hall bar edges.
The line noise was removed 
by 2D Fourier filtering.
(b) Positions $y_{\rm{k}}$ (dots) and 
widths $W_{\rm{FWHM}}$(bars) of the $\overline{\Delta V}$ profile 
as a function of $\nu$. 
Gray area: the calculated incompressible region for $\nu_{\rm{L}}=1$. 
\revisionHa{(c) Longitudinal ($R_{\rm{xx}}$) 
and normalized Hall ($\Delta R_{\rm{xy}}$) resistance curves 
measured for the same Hall bar 
at the equilibrium QH condition, $I_{\rm{sd}}=10$ nA,
at $B=8$ T. 
Vertical dashed lines mark the $\nu$ positions 
at which the SGM images shown in (a) were captured.}
}
\label{nu1map}
\end{figure}

The same technique was further applied 
to the spin-gap \revisionHa{incompressible region} 
emerging at odd $\nu_{\rm{L}}$.
Figure \ref{nu1map}(a) displays SGM images captured 
near $\nu=1$ at $B=8$ T.
The $\nu_{\rm{L}}=1$ incompressible strip was
observed as a straight line that
moves from the higher-$\mu_{\rm{chem}}$ edge (at $\nu=1.090$) 
to the interior of the Hall bar (at $\nu=1.022$).
As seen in Fig. \ref{nu1map}(b),
the measured $y_{\rm{k}}$ shifts, and $W_{\rm{FWHM}}$ widens 
with decreasing $\nu$, which is again consistent 
with the incompressible region \revisionHb{(gray area)} evaluated
with the LSDA calculations that considered
the exchange enhancement of the spin gap.

\begin{figure}
\includegraphics[width=0.9\linewidth]{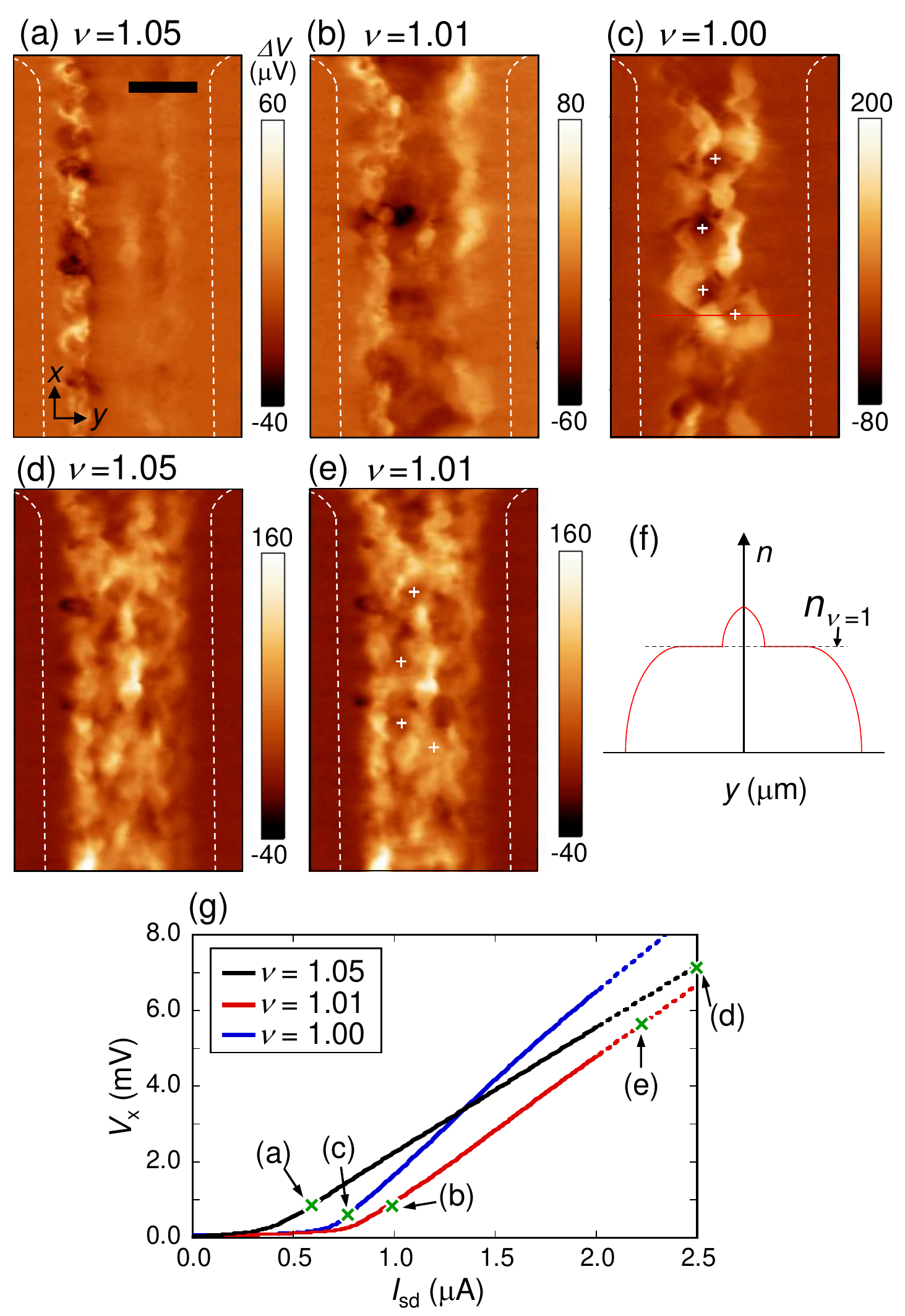}
\caption{\revisionHa{Filling factor and} 
electrical current evolution of  
\revisionHa{SGM patterns taken near $\nu=1$}
at $B=6$ T \revisionHb{(see the Supplementary Material \cite{Supple} 
for corresponding  $R_{\rm{xx}}$ and $\Delta R_{\rm{xy}}$ at equilibrium in the QH regime)}.
\revisionHa{(a)--(c) Filling-factor-dependent structures
of the $\nu_{\rm{L}}=1$ incompressible phase}
taken at low $I_{\rm{sd}}$ (0.6--1.0 $\mu$A) 
for $\nu=1.05$, 1.01, and 1.00.
\revisionHa{(d), (e) SGM patterns
obtained} at high $I_{\rm{sd}}$ (2.2--2.5 $\mu$A)  
for  $\nu=1.05$ and 1.01. 
The scale bar is 4 $\mu$m;
dashed lines denote the Hall bar edges.
The line noise was removed 
\revisionHa{using} 2D Fourier filtering.
(f) Schematic density profile along a red line 
across a closed-loop pattern in (c). 
(g) $V_{\rm{x}}$--$I_{\rm{sd}}$ curves 
for $\nu=1.05$, 1.01, and 1.00.
Crosses mark the measurement conditions for (a)--(e).
}
\label{CurrentInful}
\end{figure}

To examine the incompressible bulk \revisionHa{localization},
we focused on the \revisionHb{$\nu$ region closer to $\nu=i$} (here, $\nu=1$)
in which 
the incompressible region is expected 
to extend over the entire bulk region, 
as expected by the LSDA calculation and as shown 
in the gray region in Fig. \ref{nu1map}(b).
\revisionHb{At} $\nu=1.015$ and 1.008 \revisionHb{(Fig. \ref{nu1map}(a))},
we found a closed-loop pattern in the incompressible region.
\revisionHb{Notably, this $\nu$ region corresponds to
that in which the global longitudinal 
and Hall 
resistances at equilibrium in the QH regime
(Fig. \ref{nu1map}(c)) exhibit a dip and plateau, respectively.}
The same tendency was also observed 
in a wider spatial region 
at different $B$---namely, $B=6$ T, 
as shown in Figs. \ref{CurrentInful}(a)--(c).
In particular, Fig. \ref{CurrentInful}(c) shows 
distinct closed loop patterns (around white crosses) 
over the entire Hall bar in the $x$ direction.
The observed loop structure is attributed 
to an incompressible barrier 
encircling compressible puddles \cite{ilani2004SETprobe} 
where electrons accumulate
to screen the potential valley, as depicted 
in Fig. \ref{CurrentInful}(f) 
\revisionHa{and discussed in Supplementary Material \cite{Supple}}.
The average distance between the structures
was estimated to be about 3 $\mu$m, 
which is of the same order as  
the separation of the potential-disorder-related states
(a few $\mu$m) observed in a similar modulation-doped 
quantum well \cite{Hayakawa2013}.

\revisionHb{
The observed $\nu$ evolution of the incompressible phases
that shows agreement with the LSDA calculation
indicates that the QH phases
is microscopically robust against the local breakdown 
caused by inter-LL scattering under the nonequilibrium condition. 
The robustness of a QH state is supported
by our previously reported scanning nuclear resonance imaging \cite{HashiSNRM},
where we demonstrated the spatial homogeneity of the fully polarized 
$\nu=1$ state maintained under similar nonequilibrium conditions 
near the onset of breakdown of the QH effect. 
We speculate that this robustness maintains at nonequiliblium $I_{\rm{sd}}$
up to the limit above which the position of the incompressible pattern
showed the $I_{\rm{sd}}$ dependence (see the Supplementary Material \cite{Supple}).}

To explore how QH systems are microscopically broken 
by a larger imposed current,
we examined the current-induced evolution 
of the patterns \revisionHb{by imposing $I_{\rm{sd}}$} up to 2--5 times higher than
those used for the low-current images 
(Figs. \ref{CurrentInful}(a)--(c)), 
as indicated by the $V_{\rm{x}}$--$I_{\rm{sd}}$ curves 
(Fig. \ref{CurrentInful}(g)).
As shown in the Supplementary Movies,
both the local incompressible patterns near the edge
(Fig. \ref{CurrentInful}(a)) 
and in the bulk region (Fig. \ref{CurrentInful}(b))
gradually expand with increasing $I_{\rm{sd}}$
to the compressible region, eventually covering
the entire region and exhibiting a dense filament pattern
independent of $\nu$ (Figs. \ref{CurrentInful}(d)--(e)).
The observed vanishing of the $\nu$ dependency in the patterns
clearly indicates a \revisionHb{global} breakdown of the $\nu$-dependent QH effect. 

Compared with the QH incompressible pattern
enclosing the compressible puddle (Fig. \ref{CurrentInful}(c)),
the observed filament pattern shows
a wider distribution and corrugates 
at a shorter length scale.
Notably, the observed filaments partially
surround the positions marked by crosses
(Fig. \ref{CurrentInful}(e)),  
which correspond to the positions (crosses in Figs \ref{CurrentInful}(c))
of the disorder-induced QH compressible puddles. 
These indicate that the filament pattern is correlated 
with the potential disorder whose potential slope may  
not be fully screened owing to less compressibility
induced by the heating effect
\cite{screeningMachida1996,screeningKato2009}
in the dissipative QH breakdown regime \cite{Komiyama1996,HashiSNRM}.
This implies that inter-LL scattering arises 
along the potential disorder \cite{disorde-assistedBreakdownGuven2002} 
over the sample in the deep QH breakdown regime.

In conclusion, 
using a nonequilibrium-transport-assisted SGM technique,
we \revisionHb{demonstrated} the 
\revisionHb{robustness of the microscopic structures 
of the incompressible QH phases contributing to
topologically quantized and} nondissipative transport
in a Hall bar.
In the deep QH breakdown regime, the observed $\nu$-dependent characteristics 
vanish and are unified into a disorder-related pattern, 
suggesting that microscopic breakdown arises 
along the potential disorder of the sample.
In our future research, we shall use this powerful 
method \revisionHa{to attain} 
a microscopic understanding of nondissipative transport,
in the fractional QH effect
and other topological edge-transport effects 
of topological insulators. 
Our 
method
can probe 
local properties of topological protection,
e.g., by imaging the backscattering sites from the helical edge channel to electron puddles \cite{backscatteringPaddle}. 
\revisionHa{As such, it can tackle topical issues, such as}
the suppression of the quantized conductance of the quantum spin Hall effect \cite{SGM_HgTe,QSHE_TMD} and the nonzero longitudinal resistance
of the anomalous quantum Hall effect \cite{QAHEbackscattering,KawamuraQAHE2017},
which can be caused by backscattering.
\revisionHb{Moreover, our method can be applied to 
research on the hydrodynamics of the QH fluid \cite{eavesTurbulence,EavesPRL} 
and extended to the hydrodynamics of Dirac fluid \cite{GrapheneTurbulent}}.

We thank K. Muraki and NTT for supplying high-quality wafers, 
K. Sato and K. Nagase for sample preparation, 
M.H. Fauzi for helpful discussion, and Y. Takahashi for figure preparation. 
K.H. and T.T. acknowledge the JSPS for financial support: 
KAKENHI 17H02728 and 18K04874, respectively. 
Y.H. acknowledges support from the JSPS 
(KAKENHI 15H05867, 15K21727, and 18H01811)\revisionHa{, and S.N.
also acknowledges the JSPS for their support (15H03673).}
K.H. and Y.H. thank Tohoku University's GP-Spin program for support.
 
\bibliographystyle{apsrev4-1}
\input{references.bbl} 
\end{document}

%% file: references.bbl
%